\documentclass[preprint,superscriptaddress,amsmath,amssymb,jcp]{revtex4-1}
\usepackage{dcolumn}
\usepackage{epsfig}
\usepackage{graphics}
\usepackage[latin1]{inputenc} 
\usepackage[T1]{fontenc}
\usepackage{bm}
\usepackage{hyperref}
\usepackage{amssymb}
\newcommand{\ddst}{false}

\begin{document}

\title{Structural, vibrational, and elastic properties of a calcium aluminosilicate glass from molecular dynamics simulations: the role of the potential}

\author{M. Bauchy}
 \email[Contact: ]{bauchy@mit.edu}
 \homepage[\\Homepage: ]{http://mathieu.bauchy.com}
 \affiliation{Concrete Sustainability Hub, Department of Civil and Environmental Engineering, Massachusetts Institute of Technology, 77 Massachusetts Avenue, Cambridge, MA 02139, United States}
 \affiliation{Department of Civil and Environmental Engineering, University of California, Los Angeles, CA 90095, United States}

\date{\today}

\begin{abstract}
We study a calcium aluminosilicate glass of composition (SiO$_2$)$_{0.60}$(Al$_2$O$_3$)$_{0.10}$(CaO)$_{0.30}$ by means of molecular dynamics. To this end, we conduct parallel simulations, following a consistent methodology, but using three different potentials. Structural and elastic properties are analyzed and compared to available experimental data. This allows assessing the respective abilities of the potentials to produce a realistic glass. We report that, although all these potentials offer a reasonable glass structure, featuring tricluster oxygen atoms, their respective vibrational and elastic predictions differ. This allows us to draw some general conclusions about the crucial role, or otherwise, of the interaction potential in silicate systems.
\end{abstract}

\maketitle

\section{Introduction}
\label{sec:intro}

Classical molecular dynamics (MD) have proved to be a useful tool in studying the properties of silicate glasses, which are not always easily accessible from experiments. However, the quality of a simulation strongly depends on that of the atom--atom interaction potential \cite{hemmati_comparison_2000}. Classical potentials usually take the form of two-body, and sometimes three-body, energy terms, parameterized with respect to experimental data or \textit{ab initio} simulations. Before any further studies, it is of primary importance to check the reliability of a potential for a given system and to understand how much the obtained results depend on the potential that is used.

To better understand the effect of the potential on silicate disorder systems, we simulated a calcium aluminosilicate glass. Calcium aluminosilicate (CAS) glasses are ubiquitous in nature (e.g., magmas \cite{bauchy_viscosity_2013}) and used in industry (e.g. high-performance glasses like Gorilla $^{\circledR}$ Glass \cite{mauro_glass:_2013, mauro_unified_2012} or nuclear waste confinement glasses \cite{benoit_first-principles_2005}. Traditionally, the topology of CAS is described as a network of Si and Al tetrahedra, connected by bridging oxygen atoms (BOs) \cite{wu_evidence_1999}. On the contrary, Ca atoms depolymerize the network and create non-bridging oxygen atoms (NBOs). However, the existence of defective species, such as five-fold coordinated aluminum \cite{poe_structure_1994, stebbins_quantification_2000}, tricluster oxygen (TOs) \cite{stebbins_nmr_1997}, and free oxygen (FOs) atoms \cite{hosono_oxygen-effervescent_1987, dutt_electron_1991, dutt_structural_1992} have been reported in aluminosilicate. As these defects are not always easily accessible from experiments, it is critical to have a realistic potential to allow for microscopic MD analysis, which would lead to a better understanding of the relation between the microscopic structure and macroscopic properties.

In this paper, we report a consistent study of a calcium aluminosilicate glass using three different potentials. Structural, vibrational, and elastic properties were computed and compared with available experimental data. This allows assessing the relative quality of the different interaction models and, more generally, to better understand the role of the inter-atomic potential on the simulation of silicate systems.

\section{Potentials}
\label{sec:pot}

We aim to understand the effect of the interaction potential on computed properties of calcium aluminosilicate glasses. To this end, we selected three of the most popular potentials for CAS systems.

\textbf{The first considered potential was proposed by Matsui \cite{matsui_molecular_1996}, and has been used in several studies \cite{vargheese_origin_2010, tandia_defect-mediated_2011}. The inter-atomic interaction takes the form of a Born--Mayer--Huggins potential:}

\begin{eqnarray}
 U_{ij}(r_{ij}) = \frac{q_i q_j}{4 \pi \epsilon_0 r_{ij}} + A_{ij} \exp \left( \frac{\sigma_{ij} - r_{ij}}{\rho_{ij}} \right) - \frac{C_{ij}}{r_{ij}^6} + D_{ij}/r_{ij}^8
\end{eqnarray} \textbf{where $i$ and $j$ are atom numbers (Si, O, Al, or Ca), $r_{ij}$ is the distance between the atoms $i$ and $j$, $q_i$ is the effective charge of the atom $i$, and $A_{ij}$, $\sigma_{ij}$, $\rho_{ij}$, and $C_{ij}$  are some parameters given in Tab. \ref{tab:potq} and \ref{tab:potmat}. The three terms, respectively, represent the Coulombic, repulsive, and Van der Waals interactions. The parameters $D_{ij}$ are zero in the original version of the potential.}

\textbf{Recently, Jakse et al. reparameterized this potential \cite{bouhadja_structural_2013}, based on \textit{ab initio} calculations \cite{jakse_interplay_2012}. The refined parameters are given in Tab. \ref{tab:potq} and \ref{tab:potjak}. We chose to include this potential in the present study since, although it has the same form as Matsui's interaction, this allows us to study how small modifications of the parameters of a potential can affect the properties of the simulated system.}

\textbf{Finally, we implemented a potential proposed by Delaye \cite{delaye_investigation_2001}, and used in various studies \cite{cormier_chemical_2003, ganster_structural_2004}. The form of this potential significantly differs from that of Matsui as it features an additional higher order dipolar dispersion two-body term $D_{ij}/r_{ij}^8$ and do not rely on effective charges. The two-body parameters are given in Tab. \ref{tab:potq} and \ref{tab:potjak}. In addition, three-body interaction terms have been added to constrain the bond angles of the network forming atoms, taking the form:}

\begin{eqnarray}
 \begin{split}
  U_{ijk}(r_{ij}, r_{ik}, \theta_{ijk}) = \lambda_{ijk} \exp \left( \frac{\gamma_{ij}}{r_{ij} - r_{ij}^0} + \frac{\gamma_{ik}}{r_{ik} - r_{ik}^0} \right)  \\
  \times \left( \cos(\theta_{ijk}) - \cos(\theta_{ijk}^0) \right) ^2
 \end{split}
\end{eqnarray} \textbf{where $\theta_{ijk}$ is the angle between atoms $j$, $i$, and $k$, and $\lambda_{ijk}$, $\gamma_{ij}$, and $r_{ij}^0$ are parameters given in Tab. \ref{tab:potdel2}.}

\begin{center}
\begin{table}[h]
\caption{\label{tab:potq} Effective charges used by the three potentials \cite{matsui_molecular_1996, bouhadja_structural_2013, delaye_investigation_2001}.}
\begin{tabular}{|l|l|l|l|}
\hline
Atom & Matsui & Jakse &  Delaye \\
\hline
Si & 1.890 & 2.4 & 4.0 \\
O & -0.945 & -1.2 & -2.0 \\
Al & 1.4175 & 1.8 & 3.0 \\
Ca & 0.945 & 1.2 & 2.00 \\
\hline
\end{tabular}
\end{table}
\end{center}

\begin{center}
\begin{table}[h]
\caption{\label{tab:potmat} Two-body coefficients for Matsui's potential \cite{matsui_molecular_1996}.}
\begin{tabular}{|l|l|l|l|l|l|}
\hline
Pair & $A_{ij}$ (kcal/mol) & $\rho_{ij}$ (\AA) & $\sigma_{ij}$ (\AA) & $C_{ij}$ (kcal/mol \AA$^6$) & $D_{ij}$ (kcal/mol \AA$^8$) \\
\hline
O--O & 0.275993376 & 0.276 & 3.643 & 1962.231 & 0.0 \\
O--Si & 0.16099613 & 0.161 & 2.5419 & 1067.63 & 0.0 \\
O--Al & 0.17199587 & 0.172 & 2.6067 & 797.366 & 0.0 \\
O--Ca & 0.1779957 & 0.178 & 2.9935 & 974.51 & 0.0 \\
Si--Si & 0.04599889 & 0.046 & 1.4408 & 580.887 & 0.0 \\
Si--Al & 0.0569986 & 0.057 & 1.5056 & 433.839 & 0.0 \\
Si--Ca & 0.062998 & 0.063 & 1.8924 & 530.221 & 0.0 \\
Al--Al & 0.067998368 & 0.068 & 1.5704 & 324.01526 & 0.0 \\
Al--Ca & 0.0739982 & 0.074 & 1.9572 & 395.9991 & 0.0 \\
Ca--Ca & 0.079998 & 0.08 & 2.344 & 483.975 & 0.0 \\
\hline
\end{tabular}
\end{table}
\end{center}

\begin{center}
\begin{table}[h]
\caption{\label{tab:potjak} Two-body coefficients for Jakse's potential \cite{bouhadja_structural_2013}.}
\begin{tabular}{|l|l|l|l|l|l|}
\hline
Pair & $A_{ij}$ (kcal/mol) & $\rho_{ij}$ (\AA) & $\sigma_{ij}$ (\AA) & $C_{ij}$ (kcal/mol \AA$^6$) & $D_{ij}$ (kcal/mol \AA$^8$) \\
\hline
O--O & 0.276344 & 0.2630 & 3.6430 & 1959.372 & 0.0 \\
O--Si & 0.16120 & 0.1560 & 2.5419 & 1066.0667 & 0.0 \\
O--Al & 0.172715 & 0.1640 & 2.6067 & 796.2097 & 0.0 \\
O--Ca & 0.17732 & 0.1780 & 2.9935 & 973.0907 & 0.0 \\
Si--Si & 0.0276344 & 0.0460 & 1.4408 & 580.030 & 0.0 \\
Si--Al & 0.0575717 & 0.0570 & 1.5056 & 433.2063 & 0.0 \\
Si--Ca & 0.062177 & 0.0630 & 1.8924 & 529.445489 & 0.0 \\
Al--Al & 0.066783 & 0.0680 & 1.5704 & 323.548 & 0.0 \\
Al--Ca & 0.073691778 & 0.0740 & 1.9572 & 395.425476 & 0.0 \\
Ca--Ca & 0.080600 & 0.0800 & 2.3440 & 483.27068 & 0.0 \\
\hline
\end{tabular}
\end{table}
\end{center}

\begin{center}
\begin{table}[h]
\caption{\label{tab:potdel} Two-body coefficients for Delaye's potential \cite{delaye_investigation_2001}.}
\begin{tabular}{|l|l|l|l|l|l|}
\hline
Pair & $A_{ij}$ (kcal/mol) & $\rho_{ij}$ (\AA) & $\sigma_{ij}$ (\AA) & $C_{ij}$ (kcal/mol \AA$^6$) & $D_{ij}$ (kcal/mol \AA$^8$) \\
\hline
O--O & 8503.78796 & 0.35 & 0.0 & 0.0 & 0.0 \\
O--Si & 24063.286 & 0.328 & 0.0 & 0.0 & 0.0 \\
O--Al & 39725.5496 & 0.29 & 0.0 & 0.0 & 0.0 \\
O--Ca & 206640.707 & 0.29 & 0.0 & 12434.2219 & 20362.2376 \\
Si--Si & 20171.0765 & 0.29 & 0.0 & 0.0 & 0.0 \\
Si--Al & 22023.7562 & 0.29 & 0.0 & 0.0 & 0.0 \\
Si--Ca & 92123.7820 & 0.29 & 0.0 & 0.0 & 0.0 \\
Al--Al & 23939.6780 & 0.29 & 0.0 & 0.0 & 0.0 \\
Al--Ca & 99626.4911 & 0.29 & 0.0 & 0.0 & 0.0 \\
Ca--Ca & 412145.949 & 0.29 & 0.0 & 0.0 & 0.0 \\
\hline
\end{tabular}
\end{table}
\end{center}
  
\begin{center}
\begin{table}[h]
\caption{\label{tab:potdel2} Three-body coefficients for Delaye's potential \cite{delaye_investigation_2001}.}
\begin{tabular}{|l|l|l|l|l|l|l|}
\hline
Triplet & $\lambda_{ijk}$ (kcal/mol) & $\gamma_{ij}$ (\AA) & $\gamma_{ik}$ (\AA) & $r_{ij}^0$ (\AA) & $r_{ik}^0$ (\AA) & $\theta_{ijk}^0$ ($^o$) \\
\hline
O--Si--O & 3449.52146 & 2.6 & 2.6 & 3.0 & 3.0 & 109.5 \\
O--Al--O & 3449.52146 & 2.6 & 2.6 & 3.0 & 3.0 & 109.5 \\
Si--O--Si & 143.730061 & 2.0 & 2.0 & 2.6 & 2.6 & 160.0 \\
\hline
\end{tabular}
\end{table}
\end{center}

\section{Glass preparation}
\label{sec:glass}

We chose to study the composition (SiO$_2$)$_{0.60}$(Al$_2$O$_3$)$_{0.10}$(CaO)$_{0.30}$ as its structure can be compared with neutron diffraction data \cite{cormier_structure_2000, ganster_structural_2004}. To study the influence of the used potential, we followed a consistent approach for each glass formed. All simulations were performed with the LAMMPS package \cite{plimpton_fast_1995}, using an integration time-step of 1 fs. Coulomb interactions were evaluated by the Ewald summation method, with a cutoff of 12 \AA. The short-range interaction cutoff was chosen at 8.0 \AA. \textbf{Although they can play a critical role, we note that the values of the cutoff that are used are often omitted in publications. Here, we computed the energy of the liquid at 5000 K with respect to the chosen cutoffs and, for efficiency, picked the smallest values at which no significant evolution of the energy is observed any more.}

Liquids made of 2995 atoms were first generated by placing the atoms randomly in the simulation box. The system was then equilibrated at 5000 K in the NPT ensemble  (constant pressure) for 1 ns, at zero pressure, to assure the loss of the memory of the initial configuration. Glasses were formed by linear cooling of the liquids from 5000 to 300 K with a cooling rate of 1 K/ps. Note that, for a statistical average, we performed six independent quenchings, starting from uncorrelated liquid configurations. Once formed, glasses were relaxed at zero pressure and 300 K for 1 ns in the NPT ensemble. Subsequently, we ran 150 ps simulations in the canonical NVT ensemble for statistical averaging. In all the following, results are given at 300 K and zero pressure.

\begin{center}
\begin{table}[h]
\caption{\label{tab:rho} Densities and box length of the obtained glasses, compared with experimental densities \cite{bansal_handbook_1986, huang_structure_1991}.}
\begin{tabular}{|l|l|l|}
\hline
Potential & Density (g/cm$^3$) & Box length (\AA) \\
\hline
Matsui & 2.83$\pm$0.01 & 33.74$\pm$0.04\\
Jakse & 2.62 $\pm$0.02& 34.59$\pm$0.05\\
Delaye & 2.33$\pm$0.01 & 35.94$\pm$0.04\\
Experiment \cite{bansal_handbook_1986, huang_structure_1991, eagan_effect_1978} & 2.55--2.66 & \\
\hline
\end{tabular}
\end{table}
\end{center}

The densities of the obtained glasses are given in Tab. \ref{tab:rho} and compared with experimental values \cite{bansal_handbook_1986, huang_structure_1991}. We note that the densities largely differ from each other, which highlights the critical role of the potential. Delaye's potential tends to underestimate the density, which usually arises from the high cooling rates used in simulations \cite{bauchy_atomic_2011}. More surprisingly, Matsui's potential overestimates the density. Jakse's potential offers the best agreement with experiment, although the influence of the cooling rate should be checked.

\section{Structural results}
\label{sec:struc}

\subsection{Neutron structure factor}
\label{sec:oxy}

\begin{figure*}
\includegraphics*[height=\linewidth, angle=-90, keepaspectratio=true, draft=\ddst]{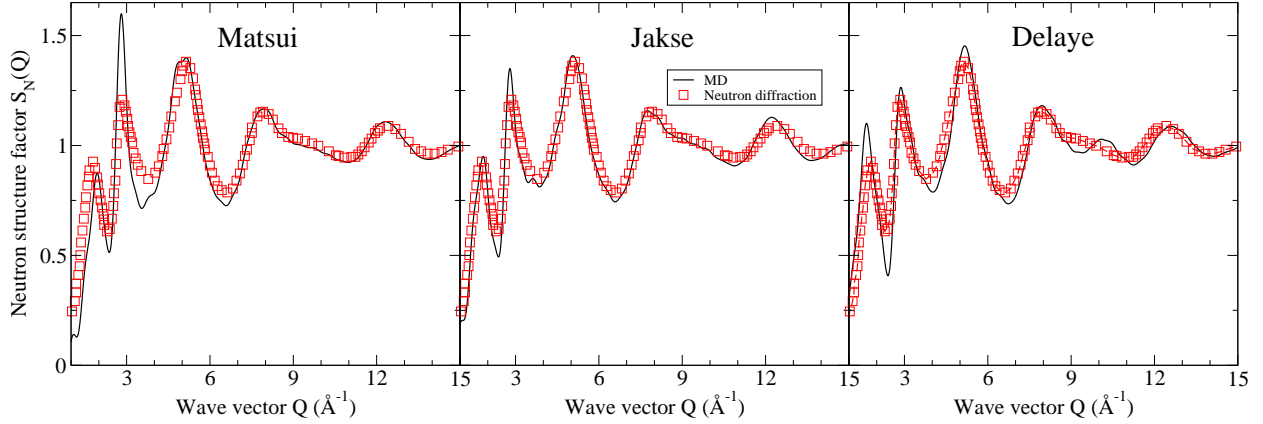}
\caption{\label{fig:Sn} (Color online) Neutron structure factors predicted by the three potentials, compared with results from neutron diffraction \cite{cormier_structure_2000, ganster_structural_2004}.}
\end{figure*}

To investigate the structure of the glass on intermediate length scales and compare with data obtained from diffraction \cite{cormier_structure_2000, ganster_structural_2004}, the neutron structure factor was computed. The partial structure factors were first calculated from the pair distribution functions (PDF) $g_{ij}(r)$:

\begin{equation}
S_{ij}(Q) = 1 + \varrho_0 \int_{0}^R 4\pi r^2 (g_{ij}(r)-1) \frac{\sin (Qr)}{Qr} F_{\text{L}}(r)\, \mathrm dr
\end{equation} where $Q$ is the scattering vector, $\varrho_0$ is the average atom number density and $R$ is the maximum value of the integration in real space (here $R$ = 16 \AA). The $F_{\text{L}}(r) = \sin (\pi r / R) / (\pi r / R)$ term is a Lorch-type window function, used to reduce the effect of the finite cutoff of $r$ in the integration \cite{wright_neutron_1988}. As discussed in Ref. \cite{du_compositional_2006}, the use of this function reduces the ripples at low $Q$, but induces a broadening of the structure factor peaks. The total neutron structure factor can then be evaluated from the partial structure factors following:

\begin{equation}
S_N(Q) = (\sum_{i,j=1}^n c_ic_jb_ib_j)^{-1} \sum_{i,j=1}^n c_ic_jb_ib_j S_{ij}(Q)
\end{equation} where $c_i$ is the fraction of $i$ atoms (Si, O, Al, or Ca) and $b_i$ is the neutron scattering length of the species (given by 4.149, 5.803, 3.449, and 4.700 fm for silicon, oxygen, aluminum, and calcium atoms, respectively \cite{sears_neutron_1992}).

Fig. \ref{fig:Sn} shows the computed neutron structure factors, each of them being compared with data from neutron scattering \cite{cormier_structure_2000, ganster_structural_2004}. We note that the experimental structure factor is fairly well reproduced by each potential, especially at high $Q$. This is not surprising, as the local structure usually weakly depends on the details of the potential. However, some differences can be observed. First, the Jakse's and Delaye's potentials provide the best reproduction of both the position and the height of the second and third peaks, even though that of Delaye predicts the existence of a small peak around 10 \AA$^{-1}$, which is not observed with other potential or in experimental data. On the contrary, Matsui's potential fails to reproduce the height of the second peak at 3 \AA$^{-1}$. The three potentials predict the existence of a first sharp diffraction peak (FSDP) around 1.7 \AA$^{-1}$, which is also observed experimentally. However, the position of the FSDP is overestimated and underestimated by Matsui's and Delaye's potentials, respectively. As the position of the FSDP is inversely proportional to a typical repetition distance in real space \cite{bauchy_structural_2012, micoulaut_anomalies_2013, bauchy_compositional_2013}, this shift is consistent with the fact that these potentials underestimate and overestimate the density, respectively. Overall, Jakse's potential provides the best agreement with neutron diffraction data.

\subsection{Radial distribution functions}
\label{sec:g}

\begin{figure*}
\includegraphics*[height=\linewidth, angle=-90, keepaspectratio=true, draft=\ddst]{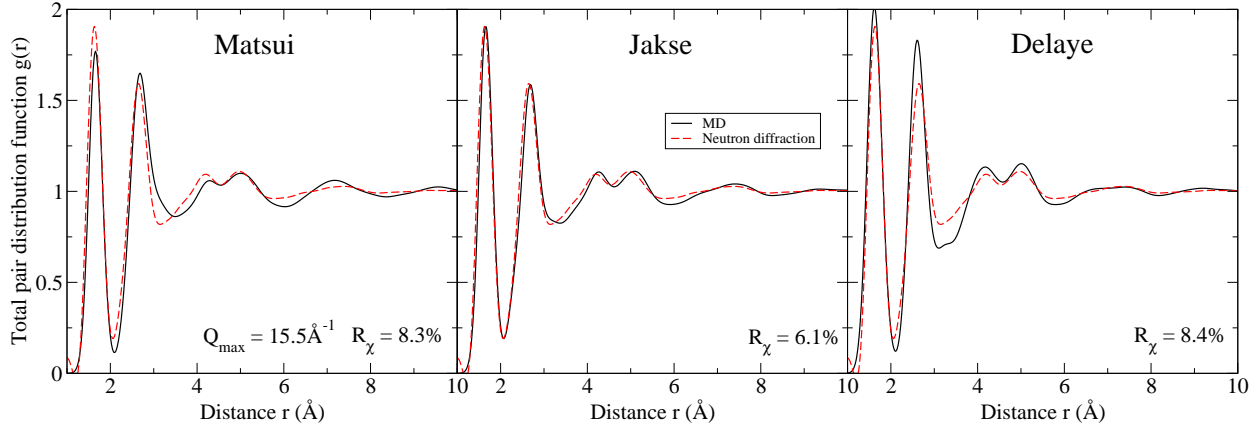}
\caption{\label{fig:g} (Color online) Total pair distribution functions predicted by the three potentials, compared with results from neutron diffraction \cite{cormier_structure_2000, ganster_structural_2004}. Respective reliability factors $R_{\chi}$ are shown for each potential.}
\end{figure*}

Since we aim to assess in detail the quality of the different potentials, we now compare their predicted structure with experimental data in real space. Indeed, as claimed by Wright \cite{wright_comparison_1993}, real space and reciprocal space correlation functions, respectively, emphasize different features of a given structure. Hence, it is necessary to compare the simulation to experiments in both spaces. Coming back to real space, the total PDFs $g(r)$ were calculated from the partials:

\begin{equation}
g(r) = (\sum_{i,j=1}^n c_ic_jb_ib_j)^{-1} \sum_{i,j=1}^n c_ic_jb_ib_j g_{ij}(r)
\end{equation} and compared to experimental data \cite{cormier_structure_2000, ganster_structural_2004}. The latter were obtained via the Fourier-transform of the experimental neutron structure factor, using the previously mentioned Lorch-type window function to reduce the ripples at low $r$. To take into account the maximal scattering vector $Q_{\rm max}$ of the experimental structure factor, the computed $g(r)$ was broadened by following the methodology described by Wright \cite{wright_comparison_1993}.

Fig. \ref{fig:g} shows the computed total PDFs for the three potentials, compared with experimental data \cite{cormier_structure_2000, ganster_structural_2004}. Once again, we observe that all three potentials offer a fair reproduction of the structure of the glass. However, the position and the height of the peaks are best reproduced by Jakse's potential. Rather than relying on a simple vidual observation, we quantified the agreement between experimental and simulated correlation functions by calculating Wright's $R_{\chi}$ factor:

\begin{equation}
R_{\chi} = \left[ \frac{\sum_{i=1}^n \left(  g(r)-g_{\rm ref}(r)\right)^2}{\sum_{i=1}^n \left( g_{\rm ref}(r)\right)^2} \right]
\end{equation} where $g_{\rm ref}(r)$ is the experimental total PDF. These factors, calculated over the range in $r$ from 1.0 \AA\ to 8.0 \AA,  are given in Fig. \ref{fig:g}. Since $R_{\chi}$ = 9 \% is typically considered as a good agreement, we conclude that the three potentials offer a realistic view of the short-range order in calcium aluminosilicate glasses. However, Jakse's potential provides the best agreement with experiments. This also means that, although convenient, relying on diffraction data might not be sufficient to discriminate among potentials.

\begin{figure*}
\includegraphics*[height=0.8\linewidth, angle=-90, keepaspectratio=true, draft=\ddst]{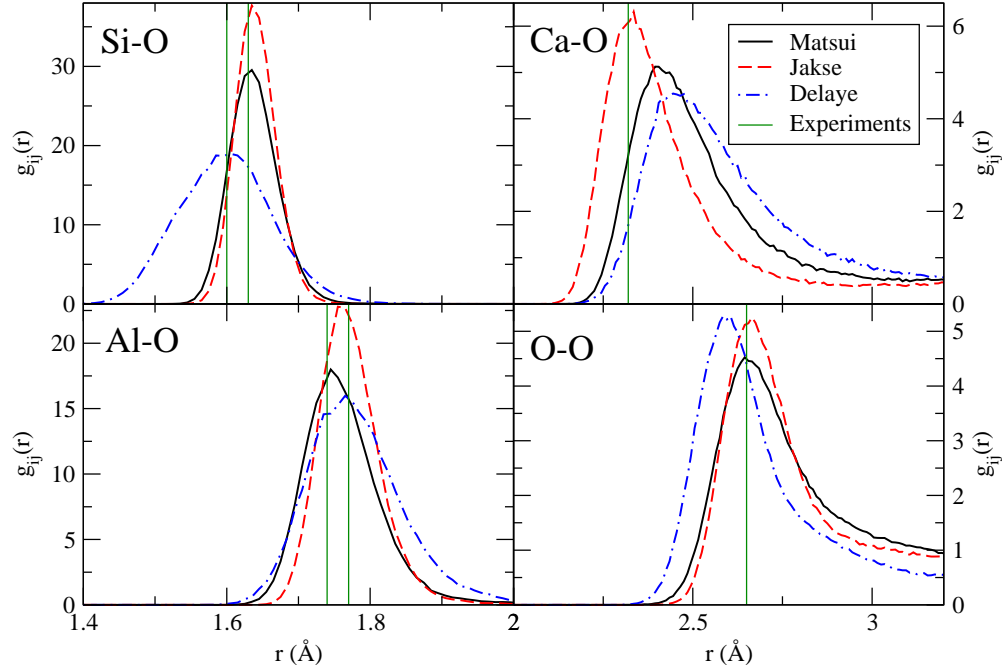}
\caption{\label{fig:gij} (Color online) Si--O, Al--O, Ca--O, and O--O partial pair distribution functions predicted by the three potentials. \textbf{Vertical lines show available experimental bond distances \cite{calas_x-ray_1987, mcmillan_raman_1982, petkov_polyhedral_2000, petkov_atomic_1998, himmel_structure_1991}.}}
\end{figure*}

\begin{figure*}
\includegraphics*[height=0.8\linewidth, angle=-90, keepaspectratio=true, draft=\ddst]{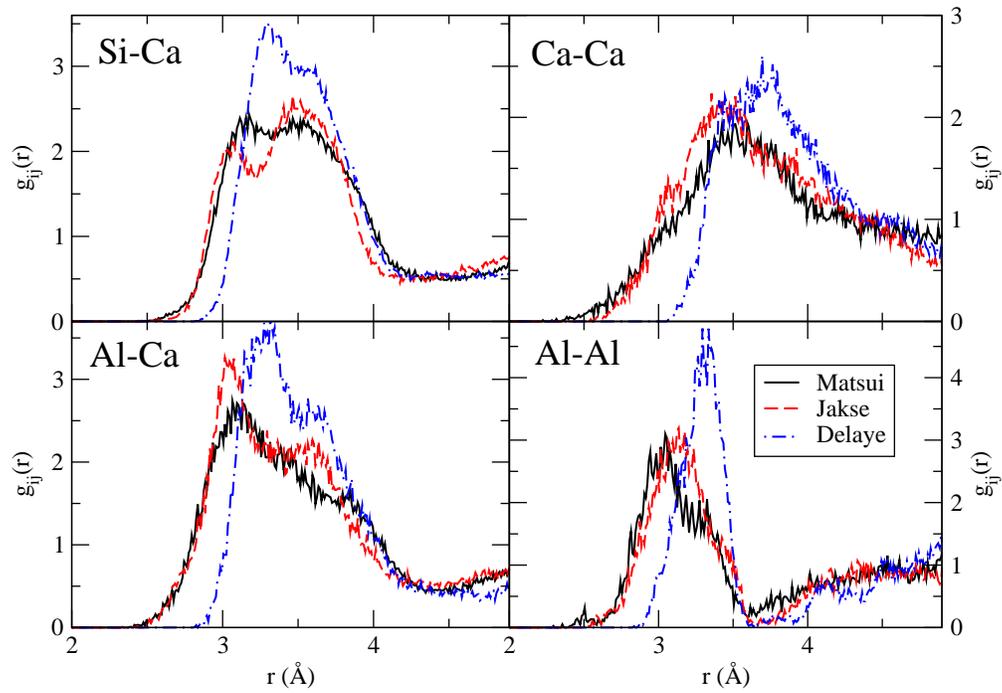}
\caption{\label{fig:gij2} (Color online) Si--Ca, Ca--Ca, Al--Ca, and Al--Al partial pair distribution functions predicted by the three potentials.}
\end{figure*}

To gain deeper insight into the local range order predicted by each potential, Figs. \ref{fig:gij} and \ref{fig:gij2} show the partial PDFs. As can be observed, although the total PDF is fairly comparable for the three potentials, the partials show larger differences, both for the position and the height of the peaks. Tab. \ref{tab:d} sums up the corresponding inter-atomic distances, compared with available experimental data. The first peak of the Si--O partial of Delaye's potential shows a broader distribution and a shift to lower $r$ with respect to the other potentials. Nevertheless, the average Si--O distance is in agreement with experiments \cite{petkov_polyhedral_2000, petkov_atomic_1998}. The Al--O partial does not show any significant change and the average position of the first peak is in good agreement with experiment \cite{mcmillan_raman_1982, calas_x-ray_1987, petkov_polyhedral_2000, petkov_atomic_1998}. On the contrary, the Ca--O partial appears to be more sensitive to the choice of the potential. Experimental values \cite{petkov_polyhedral_2000} and \textit{ab initio} simulations \cite{benoit_first-principles_2005, jakse_interplay_2012} tend to support Matsui's and Jakse's potentials for their ability to reproduce the local order around Ca atoms. The conclusion is the same for the O--O partial, as we observe a better agreement of Matsui's and Jakse's potentials with experiments \cite{himmel_structure_1991}. As observed in \textit{ab initio} simulations \cite{benoit_first-principles_2005}, the Si--Ca and Al--Ca partials show a broad first peak with a bimodal distribution with the three potentials. These bimodal distributions have been attributed to two kinds of Ca atoms, which can, respectively, be in the neighborhood of NBO or BO atoms \cite{ganster_structural_2004}. Here, and in the following, BO refers to oxygen atoms that are connected to at least two T atoms, where T = Si or Al, whereas NBO are connected to only one T atom and in the neighborhood of Ca atoms.

\begin{center}
\begin{table}[h]
\caption{\label{tab:d} Predicted interatomic distances (in \AA), compared with available experimental data \cite{mcmillan_raman_1982, calas_x-ray_1987, petkov_polyhedral_2000, petkov_atomic_1998, himmel_structure_1991}.}
\begin{tabular}{|l|l|l|l|l|}
\hline
Atomic pair & Matsui & Jakse &  Delaye & Experiment\\
\hline
Si--Si & 3.17 & 3.20 & 3.18 & 3.09 \cite{himmel_structure_1991} \\
Si--O & 1.63 & 1.63 & 1.60 & 1.60--1.63 \cite{petkov_atomic_1998, petkov_polyhedral_2000}\\
Si--Al & 3.09 & 3.19 & 3.25 & \\
Si--Ca & 3.15 & 3.07 & 3.30 & \\
Ca--Ca & 3.57 & 3.44 & 3.71 & \\
Ca--O & 2.40 & 2.32 & 2.45 & 2.32 \cite{petkov_polyhedral_2000} \\
Ca--Al & 3.11 & 3.05 & 3.27 & \\
Al--Al & 3.03 & 3.13 & 3.31 & \\
Al--O & 1.75 & 1.76 & 1.76 & 1.74--1.77 \cite{calas_x-ray_1987, mcmillan_raman_1982, petkov_polyhedral_2000, petkov_atomic_1998}\\
O--O & 2.66 & 2.66 & 2.59 & 2.65 \cite{himmel_structure_1991} \\
\hline
\end{tabular}
\end{table}
\end{center}

\subsection{Linkages}
\label{sec:link}

The Al--Al partial (see Fig. \ref{fig:gij2} is of particular interest, as it was argued that Al--O--Al linkages are energetically less favorable than Al--O--Si ones, which is known as Loewenstein's aluminum avoidance principle \cite{loewenstein_distribution_1954}. We note that the three potentials predict the existence of Al--O--Al linkages, which supports the fact that the Al avoidance principle does not necessarily hold in silicate glasses \cite{cormier_chemical_2003, ganster_structural_2004}.

\begin{center}
\begin{table}[h]
\caption{\label{tab:avoid} Number of T--O--T' linkages, where T, T' = Si or Al, compared with the prediction of a random model.}
\begin{tabular}{|l|l|l|l|l|}
\hline
Linkages & Matsui & Jakse &  Delaye & Random model \cite{ganster_structural_2004} \\
\hline
Si--O--Si & 732.3$\pm$3.5 & 747.9$\pm$2.1 & 736.0$\pm$1.5 & 1047 \\
Al--O--Al & 74.9$\pm$3.9 & 79.9$\pm$3.2 & 82.0$\pm$1.8 & 115 \\
Si--O--Al & 635.0$\pm$3.4 & 615.0$\pm$2.0 & 608.3$\pm$1.3 & 347 \\
\hline
\end{tabular}
\end{table}
\end{center}

Following the methodology presented in Ref. \cite{ganster_structural_2004}, we quantified the extent of the Al avoidance principle for the three potentials by comparing the number of T--O--T' linkages (where T, T' = Si or Al) with that predicted by a random distribution model. The results are shown in Tab. \ref{tab:avoid}. Contrary to previous simulations \cite{cormier_chemical_2003}, we clearly find an excess of Si--O--Al linkages at the expense of Si--O--Si and Al--O--Al linkages, with respect to the random distribution model predictions. This is in agreement with results from Ref. \cite{ganster_structural_2004} and suggests that the Al avoidance principle is partially satisfied in calcium aluminosilicate glasses.

\subsection{Angular distributions}
\label{sec:a}

\begin{figure*}
\includegraphics*[height=0.8\linewidth, angle=-90, keepaspectratio=true, draft=\ddst]{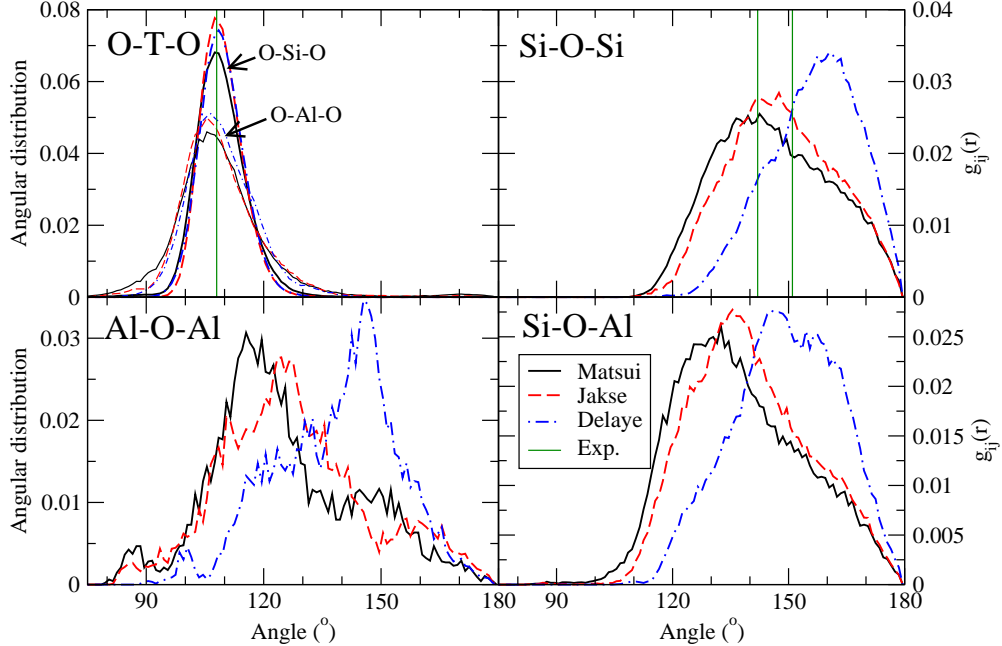}
\caption{\label{fig:angles} (Color online) O--Si--O, O--Al--O, Si--O--Si, Al--O--Al, and Si--O--Al bond angle distributions predicted by the three potentials. \textbf{Vertical lines show available experimental bond angles \cite{pettifer_nmr_1988, farnan_quantification_1992}.}}
\end{figure*}

We now focus on the bond angle distributions (BADs), which are important for understanding the extent to which the three-body potentials will improve the BAD predictions. Fig. \ref{fig:angles} shows the intra-tetrahedral O--Si--O and O--Al--O BADs, as well as inter-tetrahedral ones, Si--O--Si, Al--O--Al, and Si--O--Al. We note that the intra-tetrahedral BADs for O--T--O is fairly similar for the three potentials. The BAD for O--Si--O shows an average value of 108$^o$, in agreement with experimental results in silica \cite{pettifer_nmr_1988}. Interestingly, the O--Al--O appears to be broader and shifted to lower angle (107$^o$) with respect to the O--Si--O one, thus suggesting that Al tetrahedra are less rigid that those of Si. Intra-tetrahedral angles appear to be more sensitive to the potential and show an asymmetric shape. Overall, we observe the following trend: Si--O--Si > Si--O--Al > Al--O--Al, which is consistent with the observation that the T--O--T angle decreases with T--O distances \cite{navrotsky_tetrahedral_1985, xiao_ab_1994}. In particular, due to the use of the three-body potential, the Si--O--Si angle is narrower and centered at higher angle for the Delaye's potential, with an average of 160$^o$, compared with around 145$^o$ for the other potentials. This is a well-known issue, as classical two-body potentials, which do not include covalency or directionality in bonds,  usually fail to reproduce the value of the Si--O--Si angle in silicate glasses \cite{yuan_si-o-si_2003}. However, NMR results suggest values ranging from 142$^o$ to 151$^o$ in silica \cite{pettifer_nmr_1988, farnan_quantification_1992}. This suggests that more work is needed to calibrate the three-body terms of the Delaye's potential, as, so far, the computational cost they induce does not induce improvements of the simulated structure of the glass.

\subsection{Coordination numbers}
\label{sec:cn}

\begin{center}
\begin{table}[h]
\caption{\label{tab:cn} Predicted coordination numbers, compared with experimental data \cite{petkov_polyhedral_2000, petkov_atomic_1998, cormier_structure_2000}.}
\begin{tabular}{|l|l|l|l|l|}
\hline
Atom & Matsui & Jakse &  Delaye & Experiment\\
\hline
Si & 4.00$\pm$0.01 & 4.00$\pm$0.01 & 4.00$\pm$0.01 & 3.92 \cite{petkov_atomic_1998}, 3.95 \cite{petkov_polyhedral_2000} \\
Al & 4.08$\pm$0.06 & 4.03$\pm$0.03 & 3.96$\pm$0.04 & 4.05 \cite{petkov_atomic_1998}, 3.95 \cite{petkov_polyhedral_2000} \\
Ca & 6.9$\pm$0.2 & 6.1$\pm$0.1 & 6.9$\pm$0.2 & 5.2 \cite{petkov_atomic_1998}, 5.3 \cite{petkov_polyhedral_2000}, 7 \cite{cormier_structure_2000} \\
\hline
\end{tabular}
\end{table}
\end{center}

\begin{center}
\begin{table}[h]
\caption{\label{tab:Al} Percentage of three-, four, and five-fold coordinated Si and Al atoms.}
\begin{tabular}{|l|l|l|l|}
\hline
Species & Matsui & Jakse &  Delaye\\
\hline
Si$^{\rm IV}$ & 100 & 100 & 100 \\
Al$^{\rm III}$ & 0.0 & 0.0 & 4.1$\pm$1.3  \\
Al$^{\rm IV}$ & 93.8$\pm$1.2 & 96.8$\pm$1.0 & 95.8$\pm$1.1 \\
Al$^{\rm V}$ & 6.2$\pm$1.8 & 3.2$\pm$1.5& 0.1$\pm$0.1 \\
\hline
\end{tabular}
\end{table}
\end{center}

We now focus on the coordination numbers (CNs) predicted by the different potentials. This is of primary importance, as they strongly affect the rigidity of the network \cite{bauchy_angular_2011, bauchy_topological_2012, bauchy_transport_2013, bauchy_percolative_2013}. To evaluate the CNs, we integrated the partial PDFs up to the first minimum after the main peak. Results are shown in Tab. \ref{tab:cn}. Overall, we find that the environment of Si and Al atoms is better defined than that of Ca atoms. Hence, the CN of Ca atoms largely depends on the limit of the integration. Here, we observe that the predicted results range from 6.10 to 6.89, whereas experiments suggest values between 5.2 and 7 \cite{petkov_atomic_1998, petkov_polyhedral_2000, cormier_structure_2000}. More interesting is the case of Al atoms, as Al$^{\rm V}$ and Al$^{\rm VI}$ were found to exist in calcium aluminate liquids \cite{poe_structure_1994}. As shown in Tab. \ref{tab:Al}, we note that all potentials predict the existence of a small proportion of Al$^{\rm V}$ species. More surprisingly, Delaye's potential also features a significant amount of Al$^{\rm III}$ atoms, not observed experimentally. On the contrary, Matsui's and, to a smaller extent, Jakse's potentials tend to overestimate the fraction of Al$^{\rm V}$, which is experimentally found to be around 1\% \cite{stebbins_quantification_2000}.

\subsection{Oxygen species}
\label{sec:oxy}

\begin{center}
\begin{table}[h]
\caption{\label{tab:O} Percentage of tricluster (TO), bridging (BO), non-bridging (NBO), and free oxygen (FO) atoms, compared with predictions assuming a network of tetrahedra connected by two-fold coordinated oxygen atoms.}
\begin{tabular}{|l|l|l|l|l|}
\hline
Species & Matsui & Jakse &  Delaye & Model \\
\hline
TO & 0.9$\pm$0.2 & 0.64$\pm$0.2 & 0.48$\pm$0.08 & 0\\
BO & 76.9$\pm$0.3 & 77.0$\pm$0.1 & 76.32$\pm$0.08 &  77.73 \\
NBO & 22.1$\pm$0.2 & 22.1$\pm$0.2 &  23.20$\pm$0.08 & 22.27 \\
FO & 0.06$\pm$0.06 & 0.27$\pm$0.06 & 0.00 & 0 \\
\hline
\end{tabular}
\end{table}
\end{center}

As mentioned above, we define BOs as oxygen atoms connected to two or more T atoms, where T = Si or Al. One the contrary, NBOs are connected to only one T. If the network was simply made of tetrahedra inter-connected by two-fold coordinated oxygen atoms, then the number of NBOs would be $N_{\rm NBO}=2N_{\rm Ca}-N_{\rm Al}$ \cite{ganster_structural_2004}. At high amounts of aluminum, an excess of NBOs was observed \cite{stebbins_nmr_1997}. However, as shown in Tab. \ref{tab:O}, the computed fraction of BO and NBO do not show any significant discrepancies with this model. This contradicts the MD results for a slightly different composition using Delaye's potential \cite{ganster_structural_2004}. However, it has been reported that the percentage of NBOs decreases with the temperature \cite{jakse_interplay_2012}; hence, this contradiction can arise from the slower cooling rate used in the present study. A higher cooling rate could induce results that are more representative of the liquid phase.

\begin{center}
\begin{table}[h]
\caption{\label{tab:TO} Percentage of tricluster oxygen (TO) environments, compared with the predictions of a random network model.}
\begin{tabular}{|l|l|l|l|l|}
\hline
Environment & Matsui & Jakse &  Delaye & Model \\
\hline
OSi$_3$ & 0.00  & 0.00 & 0.00 & 0.77 \\
OSi$_2$Al & 14.1$\pm$3.2 & 0.08$\pm$0.08 & 11.2$\pm$1.8 &  10.95 \\
OSiAl$_2$ & 69.0$\pm$7.6 & 66.4$\pm$9.9 &  66.7$\pm$5.3 & 42.35 \\
OAl$_3$ & 16.9$\pm$3.7 & 33.5$\pm$9.3 & 22.1$\pm$3.5 & 45.93 \\
\hline
\end{tabular}
\end{table}
\end{center}

However, we find a small proportion of defective species (see Tab \ref{tab:O}), comprising TO atoms, i.e., tricluster O atoms, connected to three T atoms, and FO atoms, i.e., free oxygen atoms that do not show any T atom in their first coordination shell, which are typically surrounded by Ca atoms. The presence of FO atoms, although small, is surprising as they have only been observed in low silica calcium aluminosilicate glasses \cite{hosono_oxygen-effervescent_1987, dutt_electron_1991, dutt_structural_1992}. Tricluster oxygen atoms have been observed in aluminosilicate glasses \cite{stebbins_nmr_1997}. Tab. \ref{tab:TO} shows the distribution of the TBO environments, compared with the predictions of a random network distribution \cite{ganster_structural_2004}. The results clearly show an excess of OSiAl$_2$ units for the three potentials, which is in agreement with previous simulations \cite{ganster_structural_2004}. This result was interpreted as a possible charge compensation role of the oxygen triclusters \cite{ganster_structural_2004}.

\section{Vibrational results}
\label{sec:vib}

\begin{figure}
\includegraphics*[width=\linewidth, keepaspectratio=true, draft=\ddst]{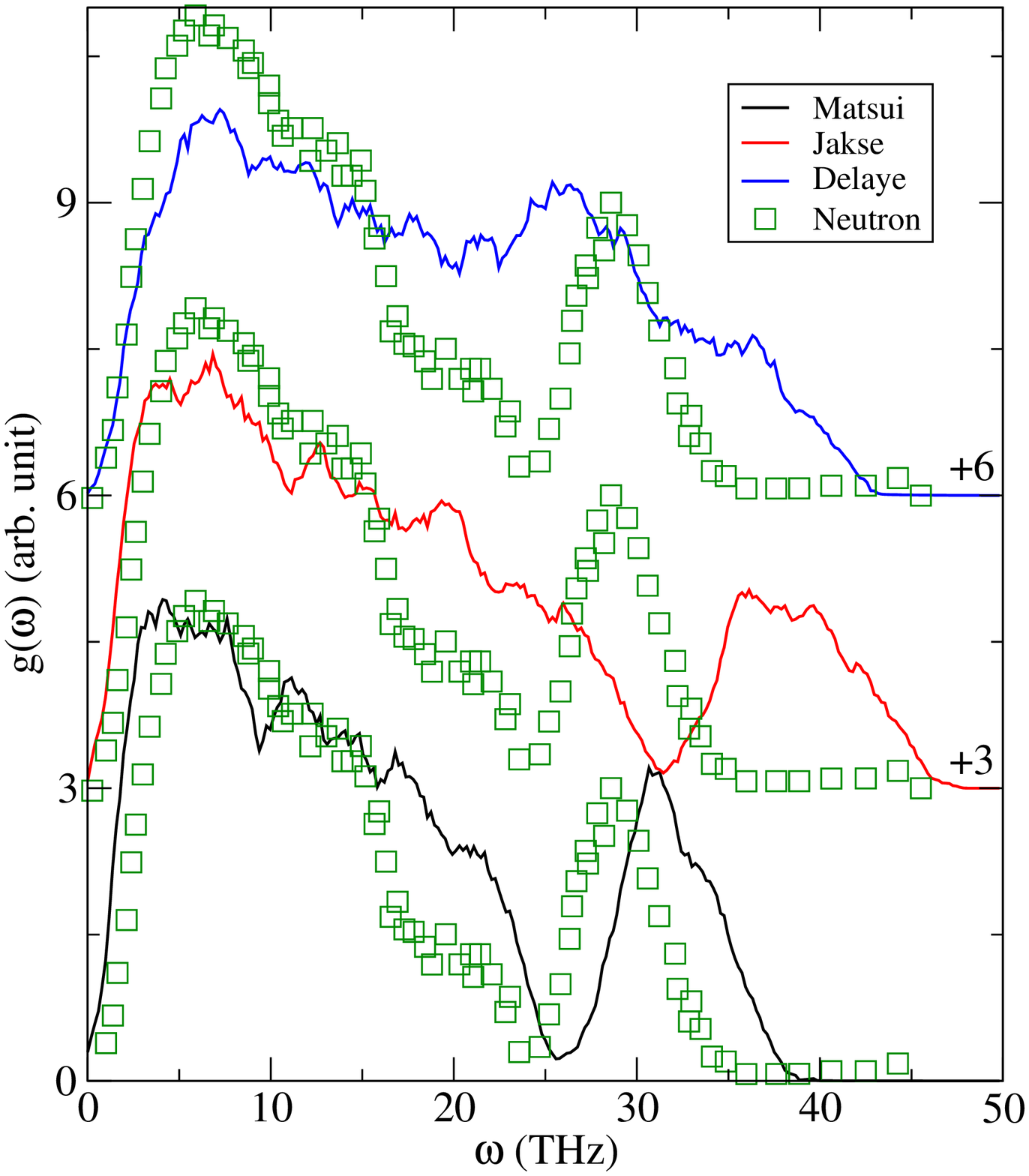}
\caption{\label{fig:vdos} (Color online) Vibrational density of states for the three potentials at $T =16$ K, \textbf{each compared with the same data from neutron scattering measurements \cite{zhao_vibrational_1997}.}}
\end{figure}

Vibrational properties are usually poorly predicted by classical potentials. We computed the vibrational density of state (VDOS) $ g(\omega)$ predicted by each potential by computing the Fourier-transform of the velocity autocorrelation function:

\begin{equation}
 g(\omega) = \frac{1}{N k_B T} \sum \limits_{j=1}^N m_j \int_{-\infty}^{\infty} <\textbf{v}_j(t) \textbf{v}_j(0)> \text{exp}(\text{i}\omega t)\, \mathrm dt
\end{equation} where $N$ is the number of atoms, $m_j$ is the mass of an atom $j$,  $\omega$ is the frequency, and $\textbf{v}_j(t)$ is the velocity of an atom $j$.

Fig. \ref{fig:vdos} shows the VDOS for each potential, computed at $T = 16$ K, compared with data from neutron measurements \cite{zhao_vibrational_1997}. Note that the experimental data are obtained for another composition [(SiO$_2$)$_{0.43}$(Al$_2$O$_3$)$_{0.14}$(CaO)$_{0.43}$]. \textbf{However, such a change of composition should not affect in a significant way the general shape of the VDOS. If the relative intensity of the peaks will obviously depend on the composition, the typical frequency of vibration should remain comparable, provided the local environmental of the atoms does not change significantly. For example, in sodium silicate, it was shown that the position of the high-frequency peak associated to Si--O stretching modes remains fairly constant with the adding of soda \cite{zotov_effects_2001}. We expect this feature to be also observed in calcium aluminosilicate glasses, but we can only rely on a qualitative comparison here.}

We note that none of the potentials offer a good reproduction of the experimental VDOS. However, Matsui's and, to a smaller extent, Jakse's potentials reproduce the general shape of the VDOS, with a main band between 0 and 25 THz and a second band, less intense, around 30 THz. These features are very similar to the VDOS of sodium silicate \cite{bauchy_structural_2012}. On the contrary, the VDOS obtained from the Delaye's potential does not show any significant gap between the low and the high frequency bands. This highlights the difficulty for classical potentials to reproduce experimental VDOS.

\section{Elasticity results}
\label{sec:elas}

\begin{center}
\begin{table}[h]
\caption{\label{tab:elas} Predicted bulk ($K$), shear ($G$), Young's moduli ($E$) and Poisson's ratio ($\nu$), compared with experimental values \cite{eagan_effect_1978}.}
\begin{tabular}{|l|l|l|l|l|}
\hline
Modulus & Matsui & Jakse &  Delaye & Experiment \cite{eagan_effect_1978}\\
\hline
$K$ & 54.7$\pm$1.3 & 64.9$\pm$1.2 & 95.6$\pm$3.0 & 77.5 \\
$G$ & 28.8$\pm$0.5 & 38.1$\pm$0.4 & 53.0$\pm$1.0 & 35.9 \\
$E$ & 73.5$\pm$1.5 & 95.6$\pm$1.4 & 134.2$\pm$3.3 & 93.3 \\
$\nu$ & 0.28$\pm$0.01 & 0.25$\pm$0.01 & 0.27$\pm$0.01 & 0.30 \\
\hline
\end{tabular}
\end{table}
\end{center}

The full stiffness tensor $C_{ij}$ was computed by calculating the curvature of the potential energy $U$ with respect to small strain deformations $\epsilon_i$ \cite{pedone_insight_2007}:

\begin{equation}
C_{ij} =  \frac{1}{V} \frac{\partial^2 U}{\partial \epsilon_i \partial \epsilon_j}
\end{equation} where $V$ is the volume of the system. We checked that the system is largely isotropic. Bulk ($K$), shear ($G$), and Young's moduli ($E$) were computed, as well as the Poisson's ratio $\nu$. These results are shown in Tab. \ref{tab:elas} and compared with experimental values for a slightly different composition \cite{eagan_effect_1978}. Similarly to the vibrational properties, elastic constants appear to be very sensitive to the choice of potential. Overall, Jakse's potential offers the best agreement with experimental values.

\section{Discussion}
\label{sec:dis}

Overall, if we restrict ourselves to the structural prediction, Jakse's potential, which results from a recalibration of the original Matsui's potential, appears to offer the best agreement with experimental data. This potential also seems to provide the best description of the mechanical properties of the glass. However, the recalibration involves an unrealistic shift in the vibrational density of states. More generally, comparing the properties predicted by different potentials allows drawing some conclusions about the effects of the potential on MD simulations of glasses.

First, we observe that all three potentials, although different in their forms, provide a realistic description of the structure of the glass, both at short- and medium-range order. This means that the generic topology of the network does not strongly depend on details of the potential; therefore, useful structural information can be obtained from MD simulations even if the potential is not perfectly calibrated. However, this study shows that potentials characterized with a reasonable structure can lead to unrealistic predictions for the VDOS and the elastic constants. Thus, if one wants to use MD to study vibrational, mechanical, or dynamical properties, comparing the predicted structure with experiments might not be sufficient to assess the ability of the potential to offer realistic values. For example, even if it was not studied here because of a lack of experimental data, diffusion and viscosity have been shown to strongly depend on the choice of potential in silicate liquids \cite{bauchy_pockets_2011, bauchy_transport_2013, bauchy_viscosity_2013}.

Second, studying the effect of different potentials allows us to better identify the features that strongly depend on details of the potential and those that do not. Hence, in the case of the present calcium aluminosilicate glass, we find a partial Al avoidance trend and the existence of Al$^{\rm V}$ and tricluster oxygen species for every potential. This suggests that these features arise from basic topological issues. On the contrary, properties that are strongly potential-dependent, like the existence of free oxygen species, are less likely to be generic, as they might arise from spurious effects of the potential.

Finally, we see that classical potentials are only approximations of the real chemical interactions between the atoms. Generally, they are good for what they have been calibrated for, but show some intrinsic limits. \textit{Ab initio} simulations offer a much more robust approach to predicting system properties. In particular, for some families of systems, like chalcogenide, they appear to be the only viable solution, as classical simulations fail to reproduce their local structures \cite{bauchy_compositional_2013, bauchy_structure_2013, bauchy_structure_2013-1, micoulaut_structure_2013}. However, first-principle simulations remain limited to small systems and short time scales, thus preventing, e.g., the study of large-scale heterogeneities or long-term relaxation. To this end, reactive potentials like REAXFF \cite{van_duin_reaxff:_2001, abdolhosseini_qomi_applying_2013, qomi_anomalous_2014, bauchy_is_2014} are an attractive approach, as they appear to be able to handle large complex systems in an accurate way while remaining faster than \textit{ab initio} simulations. However, their accuracy is still to be verified for silicate disordered materials.

\section{Conclusion}
\label{sec:ccl}

We have simulated a calcium aluminosilicate glass and studied the effects of the potential. Overall, Jakse's potential offers the best agreement with experiments for structure and elasticity, but Matsui's one provides a better prediction of vibrational properties. For the three potentials, we observe a partially satisfied aluminum avoidance effect. Moreover, the existence of tricluster oxygen atoms, primarily belonging to OSiAl$_3$ structures, is confirmed by all three potentials. Consequently, those features appear to be generic, as they do not depend on the details of the potential. More generally, this work allows us to better understand the role of the potential used in molecular dynamics studies.

\end{document}